\newcommand{\ket}[1]{\lvert #1\rangle}

\newcommand{\Jx}{J_{ex}}

\documentclass[aps,twocolumn,raggedbottom,prb,showpacs,nobalancelastpage,amssymb,groupedaddress]{revtex4}
\usepackage{graphicx}
\usepackage{amsmath}
\usepackage{amssymb}

\include{nem}
\begin{document}
\title{Perturbative regimes in central spin models}
\author{B.\, Erbe and J.\, Schliemann}
\affiliation{Institut f\"{u}r Theoretische Physik, Universit\"at
Regensburg, 93053 Regensburg, Germany}
\date{\today}
 
\begin{abstract}
Central spin models describe several types of solid state nanostructures
which are presently considered as possible building blocks of future
quantum information processing hardware\cite{Loss98}.
From a theoretical point of view, a key issue remains the treatment
of the flip-flop terms in the Hamiltonian in the presence of
a magnetic field. We consider homogeneous hyperfine and exchange coupling constants (which are different from each other) and  systematically study the influence of these terms, 
both as a function of the field strength and the size of the spin baths.
We find crucial differences between initial states with central spin
configurations of high and such of low polarizations. This has strong
implications with respect to the influence of a magnetic field on the flip-flop terms 
in central spin models of a single and more than
one central spin. Furthermore, the dependencies on bath size and
field differ from those anticipated so far. Our results might open the route 
for the systematic search for more efficient perturbative treatments of 
central spin problems.
\end{abstract}
 \maketitle

\section{Introduction}

Central spin models are the generic theoretical description for several
solid state nanostructures which are presently under intensive
experimental and theoretical study in the context of quantum information
processing. Important examples include
semiconductor \cite{Petta06, Koppens06, Hanson07, Braun05} and 
carbon nanotube \cite{Churchill09} quantum dots, phosphorus donors in 
silicon \cite{Abe04}, nitrogen vacancy centers in diamond 
\cite{Jelezko04, Childress06, Hanson08} and molecular magnets \cite{Ardavan07}. 
The typical Hamiltonian is given by
\begin{equation}
\label{H}
H = \sum_{i=1}^{N_c} A_j^{(i)} \vec{S}_i \sum_{j=1}^N \vec{I}_j + 
\sum_{i<j=1}^{N_c} J_{ij} \vec{S}_i \vec{S}_j + B\sum_{i=1}^{N_c} S_i^z
\end{equation}
and describes the interaction of $N_c$ central spins 
$\vec{S}_i$ with $N$ bath spins 
$\vec{I}_j$ characterized by coupling parameters $A_j^{(i)}$ with an overall
coupling strength $A:=(1/N_c)\sum_{i=1}^{N_c} \sum_{j=1}^N A_j^{(i)}$. 
In semiconductor quantum 
dots, for example, the role of the central spins is played by the confined electron spins
interacting with the nuclear spins of the host material via the hyperfine 
contact interaction. Here the coupling constants $A_j^{(i)}$ are proportional 
to the square modulus of the respective electronic wave function at the 
sites of the nuclear spins and therefore clearly not equal to each other 
(``inhomogeneous''). The parameters $J_{ij}$ in the second term of (\ref{H}) 
account for an exchange coupling between the different electron spins, 
where we assume $J_{ij}=:\Jx$ in the following, and the third term describes 
a magnetic field applied to the electron spins. For reviews concerning the 
hyperfine interaction in semiconductor quantum dots the reader is referred 
to Refs. \cite{SKhaLoss03,Zhang07,Klauser07,Coish09,Taylor07}.

An important ingredient to the Hamiltonian (\ref{H}) are the so-called
flip-flop terms,
\begin{equation}
H_{ff} = \frac{1}{2} \sum_{i=1}^{N_c} 
\sum_{j=1}^N A_j^{(i)} \left(S_i^+ I_j^- + S_i^- I_j^+\right)\,,
\end{equation} 
which are off-diagonal in the basis with the field direction as the
quantization axis.
Theoretical treatments of (\ref{H}) so far have usually  
distinguished between (i) the case of a strong magnetic field, as 
compared to the overall coupling strength, $B \geq A$, and (ii) the case of 
a weak magnetic field, $B < A$. In particular, for the most intensively studied
situation of a single central spin, $N_c=1$, the flip-flop terms
have in case (i) been treated as a perturbation with $A/B$ 
being a small parameter
\cite{KhaLossGla02, KhaLossGla03, Coish04, Coish05, Coish06, Coish08},
whereas in the opposite case (ii) it is commonly accepted that  
non-perturbative methods are required.
\cite{Coish09, SKhaLoss02, BorSt07, BorSt071, John09, BorSt09, BJ10} 
However, very recently it was shown in Refs. \cite{Witzel1, Witzel2}, 
again for $N_c=1$, that surprisingly there is a well-controlled perturbative 
treatment in $A(B \sqrt{N})^{-1}$, meaning that, for large enough systems, 
also the case of a \textit{weak} magnetic field can be treated perturbatively. 
This approach was motivated by the statement that the ``smallness of the longitudinal spin 
decay is controlled by the parameter $A(\Omega \sqrt{N})^{-1}$'' (see Ref. \cite{Witzel1}), 
where $\Omega$ denotes the electron spin Zeeman splitting. It is the purpose of the present paper to
give a systematic and unbiased analysis of the scaling properties regarding 
the flip-flop contributions to the dynamics and hence the perturbative regimes.

\section{Model and methods}

We want to investigate, in particular, the dependence on the bath size
$N$ so that we can not make use of exact numerical diagonalization.
For $N_c=1$ and if $A_j^{(k)}=A_{j}^{(l)}$ also for arbitrary values of $N_c$,
a natural alternative would be to choose an approach based on the
Bethe ansatz \cite{Gaudin, ErbSchlPRL}. This, however, leads to sets of
algebraic equations which are extremely difficult to treat. 
Therefore, we have to focus on the case of homogeneous couplings,
$A_j^{(i)}=A/N:=A'$ (see Refs.\cite{BorSt07, BJ10, ErbSchlPRL}). The Hamiltonian (\ref{H}) generally conserves the total spin 
$\vec{J}=\vec{S} + \vec{I}$, where 
$\vec{S}:=\sum_{i=1}^{N_c} \vec{S}_i$ and $\vec{I}:= \sum_{j=1}^N \vec{I}_j$.
For homogeneous couplings 
it in addition commutes with the square of the total bath spin $\vec{I}$,
\begin{equation}
\label{Symm}
 \left[H, \vec{J}\right]=\left[H, \vec{I}\,^2\right]=0\,.
\end{equation}
In what follows, we restrict ourselves to the spin length $S_i=I_j=1/2$. 
We calculate the central spin dynamics by decomposing the initial state
$\ket{\alpha}$ into eigenstates $\ket{\psi_i}$ of the Hamiltonian (\ref{H}),
\begin{equation}
 \ket{\alpha}= \sum_{i}\alpha_i \ket{\psi_i},
\end{equation}
and applying the time evolution operator.\cite{BorSt07, BJ10} 
We will focus on initial states with fixed $J^z$ quantum number
$m$ so that only the expectation values of the $z$-components of the
spin operators will show non-trivial dynamics. Moreover, due to the 
homogeneity of the couplings, the dynamics of the different central spins can 
be read off from each other and we therefore concentrate on the time
evolution $\langle S^z_1(t) \rangle$.

A state which is a simple product of spin states with definite 
$z$-component is, for vanishing $\Jx$, an eigenstate of the Hamiltonian
(\ref{H}) except for the flip-flop terms. Thus, for such an initial
state all dynamics is due to $H_{ff}$. Therefore, in order to isolate
the effect of the flip-flop terms we consider initial states of 
this type,
\begin{equation}
\ket{\alpha}=\ket{\underbrace{\Downarrow 
\ldots \Downarrow}_{N_c^D} \Uparrow \ldots \Uparrow} 
\otimes \ket{\underbrace{\downarrow 
\ldots \downarrow}_{N_b^D} \uparrow \ldots \uparrow}\,,
\end{equation}
so that 
\begin{equation}
 m=\frac{N_c}{2}+\frac{N}{2}-N_c^D-N_b^D.
\end{equation}
Note that for homogeneous couplings the order of the spin states
within the two subsystems is of no importance.

Since the $2^N$ dimensional bath Hilbert space is spanned by the 
eigenstates of $\vec{I}\,^2$, every product state can be written in terms of 
these eigenstates:
\begin{small}
\begin{equation}
\label{iniDec}
\ket{\underbrace{\downarrow \ldots \downarrow}_{N_b^D} \uparrow \ldots \uparrow}
= \sum_{k=0}^{N_b^D} \sum_{\left\{S_i\right\}} c_k^{\left\{S_i\right\}} 
\ket{\underbrace{\frac{N}{2}-k}_{I},\underbrace{\frac{N}{2}-N_b^D}_{m+N_c/2+N_c^D},\left\{S_i\right\}}\,
\end{equation}
\end{small}
Here the quantum numbers $\lbrace S_i \rbrace$ describe a certain 
Clebsch-Gordan decomposition of the bath. Because of (\ref{Symm}), the 
Hamiltonian (\ref{H}) does not couple states from different multiplets so that 
$\langle S_1^z(t) \rangle$ decomposes into a sum of dynamics on the 
multiplets given in (\ref{iniDec}). These contributions are weighted by 
\cite{BorSt07, BJ10}
\begin{eqnarray}
\label{dk}
\nonumber d_k&=&\sum_{\lbrace S_i \rbrace}
\left(  c_k^{\lbrace S_i \rbrace}\right)^2\\
 &=&\frac{N_b^D!(N-N_b^D)!(N-2k+1)}{(N-k+1)!k!}
\end{eqnarray}
with $k=0, \ldots, N_b^D$. In order to compute the dynamics, one still needs to
perform a diagonalization within the $2^{N_c}\times2^{N_c}$ dimensional
Hilbert space of the central spins for any value of $k$ in (\ref{iniDec}). This 
diagonalization as well as the sum according to (\ref{iniDec}) are performed numerically.

In the following we focus on a weak but finite
exchange coupling $\Jx = (1/800)A$ if not stated otherwise.
The precise value of $\Jx$ is not of
significance; similar exchange couplings of the same order
yield qualitatively the same results.
However, below we will also briefly comment on the special case  
$\Jx =0$. The polarization of the central spin system or the bath
respectively is defined by
\begin{eqnarray}
 p_{c}= \Big\lvert \frac{N_c-2N_c^D}{N_c} \Big\lvert \\
p_b=\Big\lvert \frac{N-2N_b^D}{N} \Big\lvert
\end{eqnarray}
It is well-known that the bath polarization influences the central spin
dynamics in a way very similar to a magnetic field \cite{SKhaLoss02, SKhaLoss03, BJ10}.
In the present paper, we will restrict our discussions 
to a very low bath polarization of $p_b=(1/N)$, corresponding to $N=2N_b^D+1$.
This is a particularly interesting special case because any effects of the polarization 
are excluded. However, the results to be presented below are clearly not generic with respect
to other values of $p_b$.

\begin{figure}
\begin{flushright}
\resizebox{\linewidth}{!}{\includegraphics{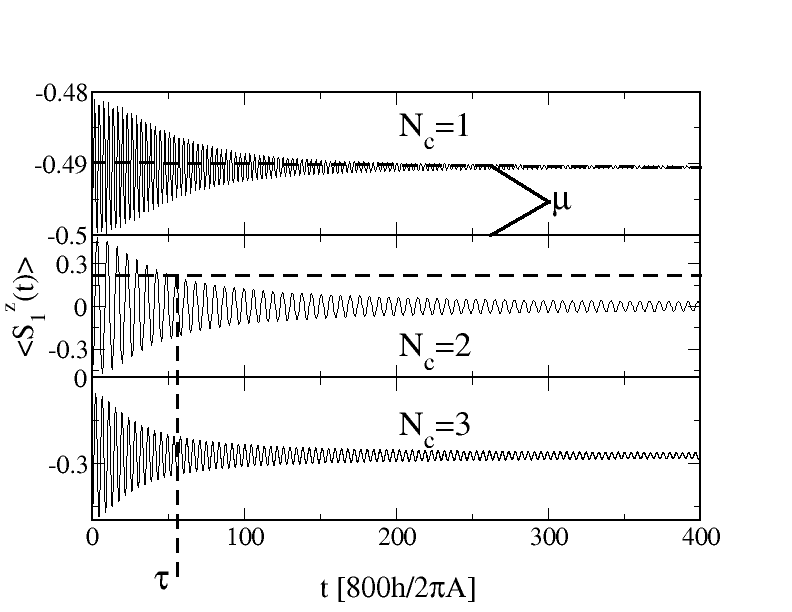}}
\end{flushright}
\caption{\label{Fig:dyn} Spin dynamics for $N_c=1,2,3$ and $N=401$. We choose an exchange coupling of $\Jx=(1/800)A$. The magnetic field is fixed to $A/B=4$. We consider initial product states with $\ket{\alpha_c}=\ket{\Downarrow},\ket{\Downarrow \Uparrow},\ket{\Downarrow \Downarrow \Uparrow}$ and a very low bath polarization of $p_b=(1/N)$. In all cases, $N_c=1,2,3$, the amplitude of the oscillation is decaying to zero (followed by a series of revivals on longer time scales not shown). The influence of the magnetic field manifests itself in a decrease of the magnitude of the spin decay and an increase of the decoherence time as measured by the decay of the amplitude of the oscillation. The two quantities, denoted by $\mu$ and $\tau$, are depicted in the first or the second panel respectively. We investigate the scaling of the decoherence time by analyzing from which time on the amplitude of the oscillation falls under a threshold level. The concrete value is of no importance, as long as the amplitude is larger than the threshold level before the onset of the decay. For high values of $p_c$ the quantity $\tau$ remains unaffected by the magnetic field and $\mu$ is the relevant perturbative measure, whereas for small $p_c$ it is $\tau$ which quantifies the influence of the flip-flop terms. }
\end{figure}

\section{The perturbative measures}

In Fig. \ref{Fig:dyn} we give examples of the dynamics for $N_c=1,2,3$ with the initial states of the central spin system, $\ket{\alpha_c}$, given by $\ket{\alpha_c}=\ket{\Downarrow}, \ket{\Downarrow \Uparrow}, \ket{\Downarrow \Downarrow \Uparrow}$. In all cases, the amplitude of the oscillation is decaying to zero (followed by a series of revivals on longer time scales not shown in the figures \cite{BJ10}). The influence of the magnetic field on the flip-flop terms manifests itself in two effects so that there are two different ``perturbative measures''. On the one hand, the spin is fixed in its initial direction, in Fig. \ref{Fig:dyn} given by $\langle S_1^z(0) \rangle=-0.5$. This means that the magnitude (``smallness'') of the spin decay, denoted by $\mu$ from now on, is decreasing with increasing magnetic field. Let us further clearify to what extent quantifications of these two effects allow to judge about the applicability of perturbative treatments. As explained above, perturbative treatments of central spin problems typically consider a magnetic field and treat $H_{ff}$ as a small perturbation. Clearly, this approach is justified only if the influence of $H_{ff}$ on the central spin dynamics is indeed sufficiently small. In other words, a perturbative treatment becomes the more adequate the stronger $H_{ff}$ is suppressed.

The quantity $\mu$ can be calculated as
\begin{equation}
\mu=\left|-0.5 - (1/T) \int_0^T dt \langle S_1^z (t) \rangle  \right|,
\end{equation}
which becomes independent of $T$ for $T \gg (\hbar/A)$. The measure is illustrated in the first panel of Fig. \ref{Fig:dyn}. On the other hand, the decoherence time, denoted by $\tau$, as measured by the decay of the amplitude of the oscillation, increases with increasing field strength. In order to calculate the scaling of $\tau$, we fix some adequate ``threshold level'' $\langle S_1^z \rangle $ and analyze after which time the amplitude falls under this value. The procedure is indicated in the second panel of Fig. \ref{Fig:dyn}. Note that the concrete value of the threshold level is of no importance. It only has to be chosen in a way that the amplitude of $\langle S_1^z(t) \rangle$ is larger than the threshold level before the onset of the decay. This procedure has been used already in Ref. \cite{BJ10}. Furthermore, a very similar approach has been chosen in Ref. \cite{BorSt07}. 

In Fig. \ref{Fig:dyn} we see that for magnetic fields of identical strengths, the value of $\mu$ is much smaller for $N_c=1$ and a central spin polarization of $p_c=1$ than for $N_c=2,3$ with $p_c=0, (1/3)$. Indeed, for large values of $p_c$ it is $\mu$ which adequately describes the influence of the magnetic field on the flip-flop terms and for small values it is $\tau$. In between, both of the measures are of relevance. In the present paper we exclude this case and concentrate on the three cases shown in Fig. \ref{Fig:dyn} where in case only one scale is relevant.

\section{Results and discussion}

Now we come to the central results of the present paper. In what follows, we investigate the scaling of the two measures $\mu$ and $\tau$ with the magnetic field strength for a fixed particle number and with the particle number for a fixed magnetic field. With respect to the scaling of $\mu$ we consider $N_c=1,2$ and for $\tau$ we focus on $N_c=2,3$. 

As already mentioned above, due to (\ref{Symm}) and (\ref{iniDec}), the dynamics $\langle S_1^z(t) \rangle$ decomposes into a sum of dynamics on different multiplets. In a first step, it is instructive to focus on $N_c=1$ and to consider only a single term of the sum. As to be demonstrated below, this leads to the scaling of the measure $\mu$ with the magnetic field on a fully analytical level. 

\subsection{Magnetic field scaling of $\mu$ for $N_c=1$}
As we are dealing with only a single central spin in this subsection, in what follows we drop the index in $\langle S_1^z(t) \rangle$. Let $I$ denote the quantum number of some multiplet in the sum $\langle S^z (t) \rangle$. This corresponds to the Hamiltonian 
\begin{equation}
 \label{HI}
H_I=A'\vec{S}\cdot \vec{I} +BS^z.
\end{equation}
For fixed $m$ this corresponds to a $2^{N_c} \times 2^{N_c}=2 \times 2$ matrix, which can be diagonalized easily. Here it is convenient to respresent $H_I$ with respect to the eigenbasis of the first term, $\vec{S} \cdot \vec{I}$, resulting from the well-known formula for coupling a spin of arbitrary length to a spin of length $S=1/2$ (see e.g. Ref. \cite{Schwabl})
\begin{eqnarray}
\label{zstates}
\nonumber \ket{I \pm \frac{1}{2},m}&=&c_{\pm}(m)\ket{\Uparrow}
\ket{I,m-\frac{1}{2}} \\
&\pm& c^{\mp}(m)\ket{\Downarrow}\ket{I,m+\frac{1}{2}},
\end{eqnarray}
where 
\begin{equation}
 c_{\pm}(m)=\sqrt{\frac{I\pm m+1/2}{2I+1}}.
\end{equation}
This yields the matrix
\begin{equation}\label{matrix}
H_I=\begin{large}
\begin{pmatrix}
\frac{A'I}{2}+ \frac{Bm}{x}&\sqrt{ \frac{B^2}{4}-\frac{B^2m^2}{x^2}}\\[5mm]
\sqrt{\frac{B^2 }{4}-\frac{B^2m^2}{x^2}}& -\frac{A'(I+1)}{2} - \frac{Bm}{x}\\
\end{pmatrix},
\end{large}
\end{equation}
where we introduced the shorthand notation $x=2I+1$. We denote the components of the $H_I$ eigenstates with respect to the basis $\lbrace \ket{\Uparrow}\ket{I,m-1/2},\ket{\Downarrow}\ket{I,m+1/2} \rbrace$ by $\psi_i^{(j)}$ and the corresponding eigenvalues by $E_i$. Diagonalizing (\ref{matrix}), we get
\begin{subequations}\label{HIeigst}
\begin{eqnarray}
\psi_1^{(1)}&=&\left(a_+ c^-(m)+b_+c^+(m)\right)\\
\psi_1^{(2)}&=&\left(b_+ c^-(m)-a_+c^+(m)\right)\\
\psi_2^{(1)}&=&\left(a_- c^-(m)+b_-c^+(m)\right)\\
\psi_2^{(2)}&=&\left(b_- c^-(m)-a_-c^+(m)\right)
\end{eqnarray} 
\end{subequations}
with
\begin{subequations}\label{ab}
\begin{eqnarray}
a_{\pm}&=&\frac{1}{\sqrt{1+\frac{4B^2x^2z_+^2z_-^2}{(A'x^2+4Bm\pm A'xy)^2}}}\\
b_{\pm}&=&\mp \frac{1}{a_{\pm}} \frac{Bz_+z_-}{Ay}.
\end{eqnarray}
\end{subequations}
Here we defined
\begin{equation}\label{y}
 y=\sqrt{x^2+\frac{4B^2 }{A'^2}+\frac{8Bm}{A'}}
\end{equation}
and
\begin{equation}
 z_{\pm}=\sqrt{\frac{x \pm 2m}{x}}.
\end{equation}
Considering the initial state $\ket{\alpha}=\ket{\Downarrow}\ket{I,m+1/2}$, it then follows for the central spin dynamics:
\begin{eqnarray}
\label{Eqdyn}
&& \langle S^z(t) \rangle =\\
\nonumber &&\underbrace{|\psi_1^{(2)}|^2 \frac{|\psi_1^{(1)}|^2-|\psi_1^{(2)}|^2}{2}}_{:=\mu_1^{(I)}} +\underbrace{|\psi_2^{(2)}|^2 \frac{|\psi_2^{(1)}|^2-|\psi_2^{(2)}|^2}{2}}_{:=\mu_2^{(I)}} \\
\nonumber &&+ \psi_1^{(2)} \psi_2^{(2)} \frac{\psi_1^{(1)} \psi_2^{(1)} - \psi_1^{(2)} \psi_2^{(2)}}{2} \cos\left[ \frac{(E_1 - E_2)t}{\hbar} \right]
\end{eqnarray}
We denote the measure $\mu$ corresponding to (\ref{HI}) by $\mu^{(I)}$. Obviously, the quantities $\mu_i^{(I)}$, introduced in (\ref{Eqdyn}), are related to $\mu^{(I)}$ by
\begin{equation}
\label{murel}
 \mu^{(I)}=\frac{1}{2}+\mu_1^{(I)}+\mu_2^{(I)}
\end{equation}
Inserting (\ref{ab}) in the expression for $\mu_{1,2}^{(I)}$ given in (\ref{Eqdyn}) and performing some extensive algebra (see appendix), we get
\begin{equation}
\label{expr}
 \mu_{1,2}^{(I)}=-\frac{1}{4} \mp \frac{(B/2A')+(m/2)}{y}+\frac{(x^2-4m^2)/4}{y^2}
\end{equation}
and hence
\begin{equation}
\mu^{(I)}=\frac{\left(x^2-4m^2\right)/2}{y^2}.
\end{equation}
Let us consider $B$ to be given in units of $A$ (for simplicity we denote $B=BA$). Then we have
$y^2=x^2+4BN(BN+2m)$. As mentioned above, in our analysis we focus on initial states
with nearly unpolarized baths. The measure $\mu$ is significant
only for highly polarized central spin systems. Hence, in the most important situation
of $N_c \ll N$ it follows $BN \gg 2m$ so that $y^2$ scales like $B^2$. Consequently,
we have $\mu^{(I)} \sim B^{-2}$. This result is independent of the value of 
$I$ and, hence, we arrive at
\begin{equation}
\label{Bscaling}
 \mu \sim \frac{1}{B^{2}}.
\end{equation}

\begin{figure}
\begin{flushright}
\resizebox{\linewidth}{!}{
\includegraphics{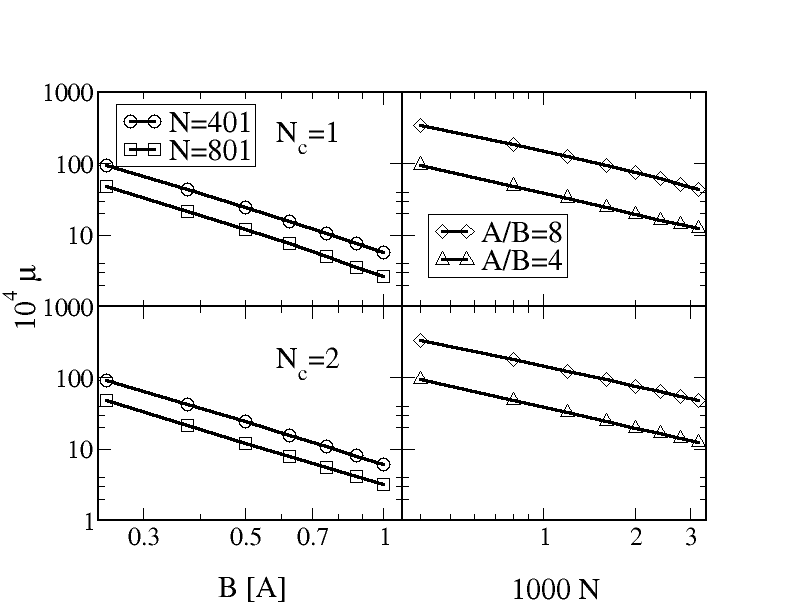}}
\end{flushright}
\caption{\label{Fig:Mu_scale} $B$ field and $N$ scalings of $\mu$ for $N_c=1,2$ and $N=401, 801$ or $A/B=8,4$, respectively. The exchange coupling in the case of $N_c=2$ is fixed to $\Jx=(1/800)A$. We consider initial product states with $\ket{\alpha_c}=\ket{\Downarrow},\ket{\Downarrow \Downarrow}$, corresponding to $p_c=1$, and a low bath polarization of $p_b=(1/N)$. The results are plotted on a double logarithmic scale. We find power laws $\sim B^{-\nu}$ with $\nu \approx 2$ and $\sim N^{-\nu}$ with $\nu \approx 1$. In the first case, the exact values are given by $\nu=2.02485, 2.09705$ ($N_c=1, N=401, 801$) and $\nu=1.95565, 1.96427$ ($N_c=2, N=401, 801$). For the $N$ scaling we have $\nu=0.989959, 1.01776$ ($N_c=1, A/B=8, 4$) and $\nu=0.925328, 0.97564$ ($N_c=2, A/B=8, 4$). With respect to the magnetic field scaling, the fully analytical result, given in (\ref{Bscaling}), is reproduced.
 }
\end{figure}

\begin{figure}
\begin{flushright}
\resizebox{\linewidth}{!}{
\includegraphics{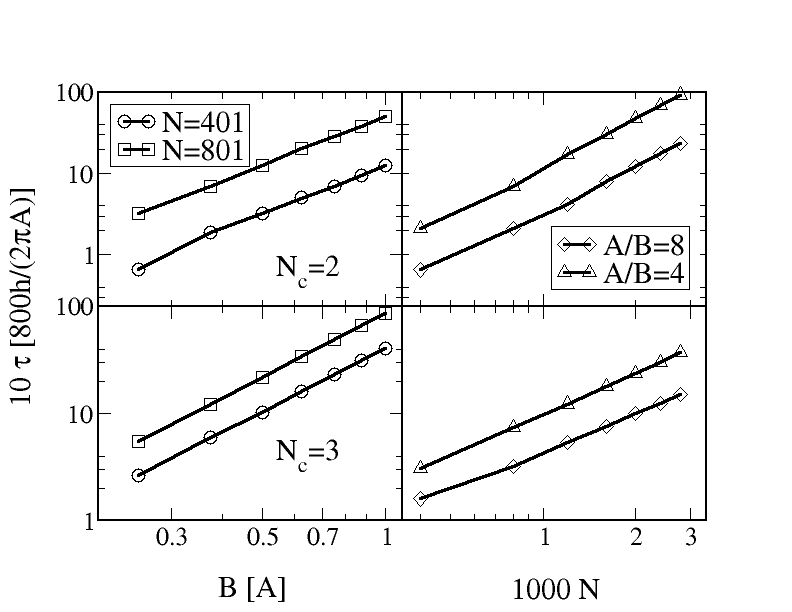}}
\end{flushright}
\caption{\label{Fig:tau_scale} $B$ field and $N$ scalings of $\tau$ for $N_c=2,3$ and $N=401, 801$ or $A/B=8,4$, respectively. The exchange coupling is fixed to $\Jx=(1/800)A$. We consider initial product states with $\ket{\alpha_c}=\ket{\Downarrow \Uparrow},\ket{\Downarrow  \Downarrow \Uparrow}$, corresponding to $p_c=0,(1/3)$, and a low bath polarization of $p_b=(1/N)$. The results are plotted on a double logarithmic scale. We find power laws $\sim B^{\nu}$ with $\nu \approx 2$ and $N^{\nu}$ with $\nu \approx 2$. In the first case, the exact values are given by $\nu=2.07975, 1.99543$ ($N_c=2, N=401, 801$) and  $\nu=1.97742, 1.99555$ ($N_c=3, N=401, 801$). For the $N$ scaling we have $\nu=1.87736,1.97722$ ($N_c=2, A/B=8, 4$) and  $\nu=1.77072, 1.92503$ ($N_c=3, A/B=8, 4$). Note that an increase of $\tau$ corresponds to a decrease of the flip-flop contributions to the dynamics. }
\end{figure}

\subsection{General scaling properties}
Now we evaluate the full dynamics in an almost analytical fashion and derive the
scalings of $\mu$ (for $N_c=1,2$) and $\tau$ (for $N_c=2,3$). As to be
demonstrated below, the number of central spins has no (direct) influence on the
result. In Figs. \ref{Fig:Mu_scale} and \ref{Fig:tau_scale} we plot $\mu$ and
$\tau$ against the magnetic field and the number of bath spins on a double
logarithmic scale. We consider two different values of $N$ or $B$ respectively
for each number of central spins. Obviously, this changes the values of the
measures, but not their scaling properties. For the $B$ field scaling we find a
simple power law $\sim B^{-2}$ in all cases, which for $N_c=1$ reproduces the
fully analytical result presented above. Indeed, already this is much stronger
than the $B^{-1}$ scaling anticipated by the perturbative approaches presented
so far. \cite{KhaLossGla02, KhaLossGla03, Coish04, Coish05, Coish06, Coish08,
Witzel1, Witzel2} Note that an \textit{increase} of $\tau$ indicates a
\textit{decrease} of the influence of the flip-flop terms on the dynamics.
The scaling with the number of baths spins turns out to be
even more surprising. For $\mu$ we find $\sim N^{-1}$, whereas for $\tau$ the
influence of $H_{ff}$ scales down with $\sim N^{-2}$. As mentioned above, for
$N_c=2,3$ we always considered a weak but non-zero exchange coupling $\Jx
=(1/800)A$. The results are generic for $\Jx \neq 0$. However, for a
\textit{zero} exchange coupling, $\Jx=0$, the $\tau$ scaling yields a slightly
different result. Here the exponent in the magnetic field scaling $\sim
B^{-\nu}$ decreases from $\nu=2$ to $\nu=(3/2)$. We have no explanation for the concrete 
value of $3/2$. However, it is not suprising that $\nu$ does not become $2$ as in the 
case $N_c=1$, where we naturally have $\Jx=0$. This would be the natural guess if for 
$\Jx=0$ the Hamiltonian would decompose in a sum of independent $N_c=1$ models. However, 
this is not the case as the $N_c$ central spins interact with a common bath. Here the dynamics 
of the central spins result from the interaction with the spin bath and with each other 
through the spin bath. The case of separate baths has been investigated in Refs.
\cite{ErbSchlEPL, ErbSchlLSA}.

Consequently, in any case the somewhat surprising approach to use $A(B \sqrt{N})^{-1}$ as a the small parameter for a perturbative treatment, presented in Refs. \cite{Witzel1, Witzel2}, turns out to even \textit{underestimate} the flip-flop suppressing character of the particle number. Hence, with respect to the particularly interesting low field case the perturbative limit is not yet achieved.

\section{Conclusion}

In conclusion, we have studied the scaling of the influence of the flip-flop terms with the magnetic field and the number of bath spins in central spin models. In order to be able to treat comparatively large systems, we considered homogeneous couplings. The flip-flop contribution to the dynamics has been isolated by choosing simple product initial states. The effect of an applied magnetic field manifests itself in the magnitude of the spin decay and the decoherence time. For highly polarized central spin systems it is the magnitude of the spin decay which is the relevant scale, whereas for a low central spin polarization it remains, to a large extent, unaffected and the decoherence time describes the influence of the magnetic field. 
We investigated the scaling of $\mu$ and $\tau$ for $N_c=1,2$ and $N_c=2,3$ in different parameter regimes. 

Surprisingly, we found that $\mu$ decreases \textit{quadratically} with the
magnetic field and linearly with the number of bath spins. For $N_c=1$ we
presented a fully analytical derivation. For $\Jx >0$ the decoherence time shows
identical scaling properties with respect to the magnetic field, whereas if
$\Jx=0$, the behavior slightly changes to $\sim B^{-3/2}$. As a very interesting
and unexpected result, it turns out that $\tau$ increases \textit{quadratically}
with the number of bath spins, corresponding to a quadratic decrease of the
flip-flop contributions to the dynamics. Summarizing, for $p_c \approx 1$ our
results suggest
\begin{equation*}
A(B \sqrt{N})^{-2}
\end{equation*}
as the small parameter of a perturbation theory. This essentially goes along with the approach considered in Refs. \cite{Witzel1, Witzel2}. However, for small values of $p_c$ we have 
\begin{equation*}
A(BN)^{-2}.
\end{equation*}
This means that the perturbative treatments of central spin models presented so far strongly underestimate the influence of both, the magnetic field as well as the number of bath spins. It is therefore desirable to search for new approaches using the full suppression of the flip-flop terms. Although all scaling properties are independent $N_c$, central spin models with more than one central spins are particularly interesting with respect to such investigations, as here low polarizations of the central spin system can be achieved. This leads to a very strong decrease of the influence of the flip-flop terms with the number of bath spins and hence the possibility to treat extremely small magnetic fields.

\section{Acknowledgments}

This work was supported by DFG via SFB631.

\section*{Appendix}

In what follows we present details on the derivation of (\ref{expr}). Inserting the eigensystem (\ref{HIeigst}) of the Hamiltonian (\ref{HI}) in the expression for $\mu^{(I)}_{1,2}$ given in (\ref{Eqdyn}) we get
\begin{widetext}
\begin{equation}\label{mu}
\mu_{1,2}^{(I)}=-\frac{z_+^2(B+A'm)\left(A'x^2+4Bm\pm A'xy\right)\left(A'x^2\pm A'xy+2B\left(2m+xz_-^2\right)\right)^2}{4x(A'^2x^3\pm A'x\left(\pm 8Bm+A'xy\right)+4B\left(B+B(x-1)\pm Amy\right))^2},
\end{equation}
\end{widetext}
which can be simplified to
\begin{widetext}
\begin{equation}\label{rewrit}
 \mu^{(I)}_{1,2}= \frac{-\left(B+A' m\right)\left(x+2m\right)\left(A' x^2+2Bx\pm A' xy\right)^2}{4x^2A'^2y^2\left(4Bm+A'x^2\pm A'xy\right)}.
\end{equation}
\end{widetext}
On the other hand, the expression (\ref{expr}) can be rewritten as
\begin{equation}\label{expr2}
 \mu^{(I)}_{1,2}=\mp \frac{(B+A'm)}{2Ay}+\frac{(B+A'm)^2}{A'^2y^2}.
\end{equation}
If we now equalize (\ref{rewrit}) with (\ref{expr2}) and multiply by the denominators, we get for
one side
\begin{eqnarray}\label{xx}
\nonumber&-& 16B^2m\mp 8A'Bmy-16A'Bm^2 \\
\nonumber&\mp& 2A'^2x^2y-4A'Bx^2-4A'^2x^2m \\
&-& 2A'^2xy^2\mp 4A'Bxy\mp 4A'^2mxy 
\end{eqnarray}
and for the other 
\begin{eqnarray}\label{xxx}
\nonumber&-&A'^2x^3\mp A'^2x^2y-2A'Bx^2 \\
\nonumber&\mp& A'^2x^2y- A'^2xy^2- 2A'Bxy\\
\nonumber&-& 2A'Bx^2\mp 2A'Bxy- 4B^2x \\
\nonumber&-& 2A'^2x^2m\mp 2A'^2xym - 4A'Bmx\\
\nonumber&\mp& 2A'^2xym- 2A'^2y^2m\mp 4A'Bmy\\
&-& 4A'Bmx\mp 4A'Bmy- 8B^2m
\end{eqnarray}
Inserting (\ref{y}) in the terms proportional to $y^2$ immediately 
shows that (\ref{xx}) and (\ref{xxx}) are identical, which
yields (\ref{expr}).


\begin{thebibliography}{99}

\bibitem{Loss98} D. Loss and D.~P. DiVincenzo, Phys. Rev. A 
\textbf{57}, 120 (1998).
\bibitem{Petta06} J.~R. Petta, A.~C. Johnson, J.~M. Taylor, E.~A. Laird, A. Yacoby, M.~D. Lukin, C.~M. Marcus, M.~P. Hanson, and A.~C. Gossard, Science
\textbf{309}, 2180 (2005).
\bibitem{Koppens06} F.~H.~L. Koppens, C. Buizert, K.~J. Tielrooij, I.~T. Vink, K.~C. Nowack, T. Meunier, L.~P. Kouwenhoven, and L.~M.~K. Vandersypen, Nature (London) \textbf{442}, 766 (2006).
\bibitem{Hanson07} R. Hanson, L.~P. Kouwenhoven, J.~R. Petta, S. Tarucha, and L.~M.~K. Vandersypen, Rev. Mod. Phys. 
\textbf{79}, 1217 (2007).
\bibitem{Braun05} P.-F. Braun, X. Marie, L. Lombez, B. Urbaszek, T. Amand, P.
Renucci, V.~K. Kalevich, K.~V. Kavokin, O. Krebs, P. Voisin, and Y. Masumoto,
Phys. Rev. Lett. \textbf{94}, 116601 (2005).
\bibitem{Churchill09} H. Churchill, A. Bestwick, J. Harlow, F. Kuemmeth, D. Marcos, C. Stwertka, S. Watson, and C. Marcus, Nat. Phys. 
\textbf{5}, 321 (2009).
\bibitem{Abe04} E. Abe, K.~M. Itoh, J. Isoya, and S. Yamasaki, Phys. Rev. B 
\textbf{70}, 033204 (2004).
\bibitem{Jelezko04} F. Jelezko, T. Gaebel, I. Popa, A. Gruber, and J. Wrachtrup, Phys. Rev. Lett. 
\textbf{92}, 076401 (2004).
\bibitem{Childress06} L. Childress, M.~V. Gurudev Dutt, J.~M. Taylor, A.~S. Zibrov, F. Jelezko, J. Wrachtrup, P.~R. Hemmer, and M.~D. Lukin, Science 
\textbf{314}, 281 (2006).
\bibitem{Hanson08} R. Hanson, V.~V. Dobrovitski, A.~E. Feiguin, O. Gywat, and D.~D. Awschalom, Science 
\textbf{320}, 352 (2008).
\bibitem{Ardavan07} A. Ardavan, O. Rival, J.~J.~L. Morton, S.~J. Blundell, A.~M. Tyryshkin, G.~A. Timco, and R.~E.~P. Winpenny, Phys. Rev. Lett. 
\textbf{98}, 057201 (2007).
\bibitem{SKhaLoss03} J. Schliemann, A.~V. Khaetskii, and D. Loss , 
J. Phys.: Condens. Mat. \textbf{15}, R1809 (2003).
\bibitem{Zhang07} W. Zhang, N. Konstantinidis, K.~A. Al-Hassanieh, and V.~V. Dobrovitski,
J. Phys.: Condens. Mat. {\bf 19}, 083202 (2007).
\bibitem{Klauser07} D. Klauser, D.~V. Bulaev, W.~A. Coish, and D. Loss, arXiv:0706.1514.
\bibitem{Coish09} W.~A. Coish and J. Baugh, phys. stat. sol. B  {\bf 246}, 2203 (2009).
\bibitem{Taylor07} J.~M. Taylor, J.~R. Petta, A.~C. Johnson, A. Yacoby, C.~M. Marcus, and M.~D. Lukin, Phys. Rev. B
\textbf{76}, 035315 (2007).
\bibitem{KhaLossGla02} A.~V. Khaetskii, D. Loss, and L. Glazman, Phys. Rev. Lett.
\textbf{88}, 186802 (2002).
\bibitem{KhaLossGla03} A.~V. Khaetskii, D. Loss, and L. Glazman, Phys. Rev. B
\textbf{67}, 195329 (2003).
\bibitem{Coish04} W.~A. Coish and D. Loss, Phys. Rev. B 
\textbf{70}, 195340 (2004).
\bibitem{Coish05} W.~A. Coish and D. Loss, Phys. Rev. B 
\textbf{72}, 125337 (2005).
\bibitem{Coish06} D. Klauser, W.~A. Coish, and D. Loss, Phys. Rev. B 
\textbf{73}, 205302 (2006).
\bibitem{Coish08} D. Klauser, W.~A. Coish, and D. Loss, Phys. Rev. B 
\textbf{78}, 205301 (2008).
\bibitem{SKhaLoss02} J. Schliemann, A.~V. Khaetskii, and D. Loss, Phys. Rev. B
\textbf{66}, 245303 (2002).
\bibitem{BorSt07} M. Bortz and J. Stolze, J. Stat. Mech. 
P06018 (2007).
\bibitem{BorSt071} M. Bortz and J. Stolze, Phys. Rev. B
\textbf{76}, 014304 (2007).
\bibitem{John09} J. Schliemann, Phys. Rev. B
\textbf{81}, 081301(R) (2010). 
\bibitem{BorSt09} M. Bortz, S. Eggert, and J. Stolze, Phys. Rev. B
\textbf{81}, 035315 (2010).
\bibitem{BJ10} B. Erbe and J. Schliemann, Phys. Rev. B 
\textbf{81}, 235324 (2010).
\bibitem{Witzel1} L. Cywinski, W.~M. Witzel, and S. Das Sarma, Phys. Rev. Lett. 
\textbf{102}, 057601 (2009).
\bibitem{Witzel2} L. Cywinski, W.~M. Witzel, and S. Das Sarma, Phys. Rev. B 
\textbf{79}, 245314 (2009).
\bibitem{Gaudin} M. Gaudin, J. Phys. (Paris)
\textbf{37}, 1087 (1976).
\bibitem{ErbSchlPRL}  B. Erbe and J. Schliemann, Phys. Rev. Lett. 
\textbf{105}, 177602 (2010).
\bibitem{Schwabl} F. Schwabl, \emph{Quantum Mechanics},
 (Springer, Berlin 2002) chapter 10.3.
\bibitem{ErbSchlEPL}  B. Erbe and J. Schliemann, Europhys. Lett. 
\textbf{95}, 47009 (2011).
\bibitem{ErbSchlLSA} B. Erbe and J. Schliemann, Phys. Rev. B 
\textbf{85}, 155127 (2012). 


\end{thebibliography}
\end{document}